\documentclass[preprint,12pt]{elsarticle}

\usepackage{natbib}

 \usepackage{graphics}
\usepackage{graphicx}
\usepackage{epsfig}
\usepackage{epstopdf}
\usepackage{amssymb}
 \usepackage{amsthm}

 \usepackage{lineno}
\usepackage{setspace}

\long\def\symbolfootnote[#1]#2{\begingroup%
\def\thefootnote{\fnsymbol{footnote}}\footnote[#1]{#2}\endgroup}

\journal{Nuclear Physics A}

\begin{document}

\begin{frontmatter}



\title{Permeability of Noble Gases through Kapton, Butyl, Nylon, and ``Silver Shield"}

\author{Steven J. Schowalter}
\author{Colin B. Connolly\symbolfootnote[1]{Corresponding author (e-mail: connolly@physics.harvard.edu)}}
\author{John M. Doyle}
 \address{Department of Physics, Harvard University, 17 Oxford St., Cambridge, MA 02138, USA}

\begin{abstract}
Noble gas permeabilities and diffusivities of Kapton, butyl, nylon, and ``Silver Shield" are measured at temperatures between $22^{\circ}$C and $115^{\circ}$C.  The breakthrough times and solubilities at $22^{\circ}$C are also determined.  The relationship of the room temperature permeabilities to the noble gas atomic radii is used to estimate radon permeability for each material studied.  For the noble gases tested, Kapton and Silver Shield have the lowest permeabilities and diffusivities, followed by nylon and butyl, respectively.
\end{abstract}


\end{frontmatter}

 \linenumbers


\section{Introduction}
The permeability of radon through the polyimide Kapton \cite{kapton} is a key factor in determining its
effectiveness as a gasket or membrane material in certain low radioactive
background experiments, such as MiniCLEAN \citep{CLEAN, MiniCLEAN}. Kapton is a
polyimide manufactured by DuPont  and has applications in aerospace design,
electrical insulation, automotive design, vacuum experiments, and
more \citep{hammound, hioki, baudouy}.  Its utility in many applications is due to its ability to retain
certain desirable properties when cooled to low temperatures, for example its pliability.  This is most dramatically shown by its use as a superfluid-tight seal gasket at temperatures below $2$ K \cite{richardson}. Also, Kapton film is
relatively inexpensive and can be easily formed, making it an
appealing material for other experimental applications.  In the MiniCLEAN experiment, Kapton is a candidate to perform a sealing
function for about one hundred roughly $25$ cm diameter flanges at temperatures
between $20$-$300$ K.  This gasket must keep radon from
permeating into the main vacuum vessel while at room temperature.

Here we report measurements of noble gas permeation through Kapton film and other technical materials including nylon \citep{nylon}, butyl \citep{butyl}, and ``Silver Shield" \citep{silvershield}, all of which have uses as gaskets, in gloveboxes, or as shielding from radon permeation.  Nylon is frequently used as a bagging material to prevent radon from coming in contact with detector components during shipping or storage.  Butyl is an inexpensive and resilient glove material and can be used as a vacuum seal gasket.  Silver Shield is a composite glove or bagging material specifically designed for low permeability that includes layers EVOH (polyvinyl alcohol), which has been shown to have low permeability to radon \cite{EVOH}.

\section{Background}\label{background}
Permeation is the process through which a gas passes through a solid material. The permeability $K$ is defined as
\begin{center}
\begin{equation}\label{eq:permrate}
Q=K\frac{A}{d}\Delta P
\end{equation}
\end{center}
where $Q$ is the number flow rate of a test gas through a thickness $d$ and cross-sectional area $A$ under a pressure difference $\Delta P$.  The permeability $K$ can also be written as 
\begin{center}
\begin{equation}\label{eq:Db}
K=Db
\end{equation}
\end{center}
where $D$ is the diffusivity and $b$ is the solubility of gas in the material.  The solubility determines the concentration of gas dissolved in the polymer at a given partial pressure; the diffusivity determines the rate at which gas flows in the material.

 By observing the time evolution of gas permeation after
establishing a concentration gradient, it is possible to probe
diffusivity independent of solubility.  The solution of the one-dimensional diffusion equation \citep{barrer} for gas diffusing across a membrane of thickness $d$ gives the gas flow $Q$ from the low-pressure side to be 
\begin{equation}\label{eq:diffeq}
Q(t)=Q_{0}[1+2\displaystyle \sum_{n=1}^{\infty} (-1)^{n}\exp{(-(n\pi)^{2}\frac{d^2}{D}t)}].
\end{equation}
where $Q_{0}$ is the final steady state flow.  Note that the dynamics of the flow are determined only by $d$ and the diffusivity $D$. The time taken for a significant amount of gas to permeate through the film is called the breakthrough time or lag time. Experiments measuring permeation typically define this to be
\begin{equation}\label{eq:lagtime}
t_{b}=\frac{d^2}{6D}.
\end{equation}
The determination of $D$ and $t_b$ from flow measurements is discussed in detail in Section \ref{resultsanddiscussion}.   

As with permeation through other polymers, the permeation of noble gases through the materials studied is expected to increase with increasing temperature. The permeability and breakthrough time are expected to follow the relations
\begin{center}
\begin{equation}\label{eq:temp_perm}
K(T) \propto  exp(-E_{K}/k_{B} T)
\end{equation}
\end{center} 
\begin{center}
\begin{equation}\label{eq:temp_tb}
t_{b}(T) \propto d^{2} exp(E_{D}/k_{B}T)
\end{equation}
\end{center}
where  $E_{K}$ is the energy of permeation, and $E_{D}$ is the energy of diffusion.   In this experiment, this temperature dependence is observed and used to extrapolate room temperature ($22^{\circ}$C) xenon permeability for Kapton.  Ultimately any temperature dependence can be exploited in order to
increase or decrease the rate of permeation.

\section{Experimental}\label{experimental}
We measure permeation using a specific gas flow method in which a constant high pressure of gas is placed on one side of a film and the steady-state pressure of permeated gas is monitored with a calibrated Residual Gas Analyzer (RGA) on the low pressure, evacuated, side of the film. Our design enables us to measure the permeability and diffusivity for helium, neon, argon, krypton, and xenon through various membrane materials. Due to the highly radioactive nature of radon, measuring the permeation of radon in this manner would be too onerous.  Instead we estimate the permeation rate of radon by extrapolating from permeation data of the stable noble gases.

\begin{figure}[p]
\begin{center}
\includegraphics[width=90mm]{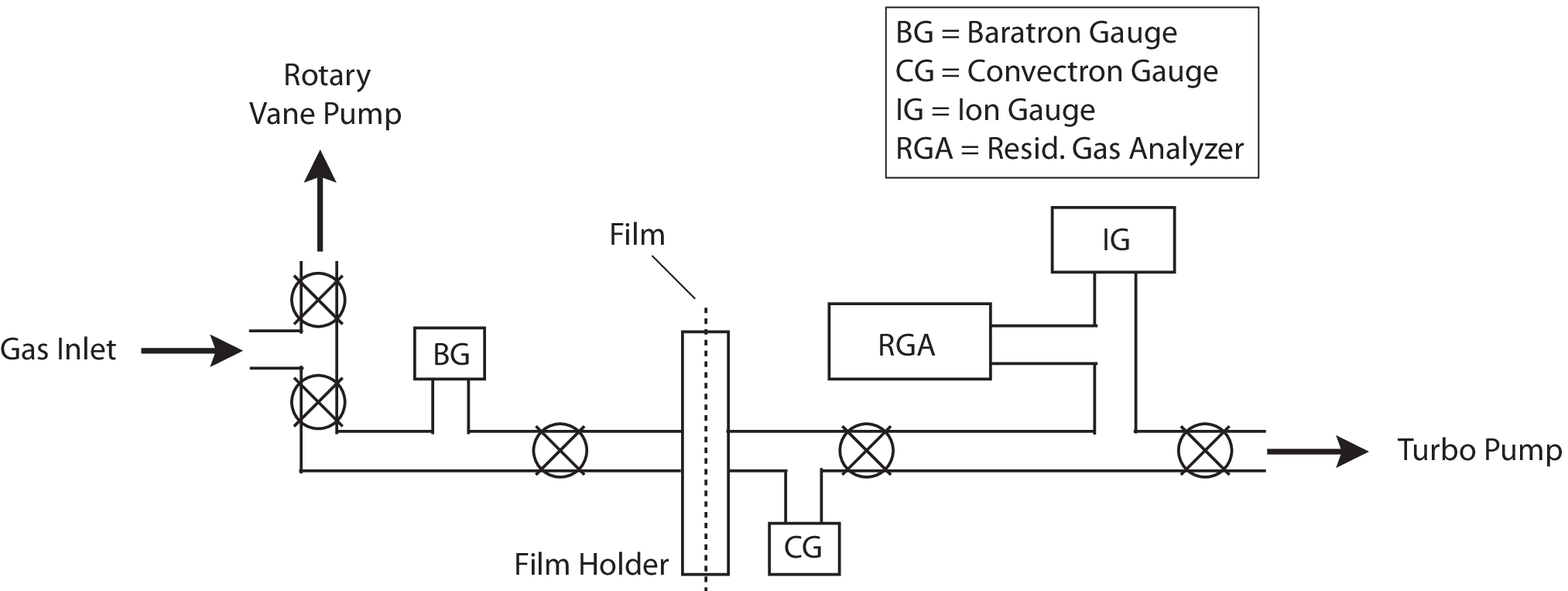}
\caption{A schematic of the apparatus used to measure permeation.}
\label{apparatus}
\end{center}
\end{figure}

The apparatus (shown in Figure \ref{apparatus})
consists of three major parts: a high-pressure inlet chamber, a low-pressure outlet
chamber, and a film holder.  

The two chambers are constructed from stainless steel tubes and VCR fittings and connected to two vacuum pumps.  The high-pressure chamber is connected to a rotary vane pump which is able to evacuate the chamber to pressures of $10^{-3}$ torr prior to filling with test gas.  A simple gas handling system introduces up to $10^{3}$ torr of test gas into the high-pressure chamber (as measured by a Baratron pressure gauge).  The low-pressure chamber is connected to a turbomolecular pump capable of evacuating the chamber to $10^{-6}$ torr, as well as to a xenon standard leak (SL), an ionization gauge, and an RGA. 

The high- and low-pressure chambers are separated by a film of the material under study housed in a film holder.  The film holder consists of two custom flanges, one made of brass and one made of aluminum, and each makes a Viton O-ring seal to one side of the Kapton film. The film is pressed between the O-rings, which are held in grooves in the flanges.  Each flange has a fitting in order to connect the film holder between the high- and low-pressure chambers. To minimize the chances of the film warping or rupturing under differential pressure (as high as $10^{3}$ torr), a depression on the inside of the low-pressure flange holds a stainless steel mesh with a grid size of $2$ mm and $40\%$ open area, which provides mechanical support for the film.  The cross sectional area for test gas diffusion is  $83$ cm$^{2}$.

To manipulate the temperature of the film, heater tape and insulation are wrapped around the metal film holder.  The temperature is monitored by thermocouples attached at various places on the film holder.

An RGA is used to measure and distinguish partial pressures of different gases below \(10^{-4}\) torr in the high-vacuum chamber. Once both experimental chambers have been evacuated, test gas (such as argon) is introduced into the high-pressure chamber to establish a pressure gradient across the film.  As the test gas begins to permeate, the RGA partial pressure rises asymptotically to a steady state value, $P_{ss}$, set by the flow of the permeating gas and by the pumping speed and conductance of the pumping line. 

\section{Results and Discussion}\label{resultsanddiscussion}

\begin{figure}[b]
\begin{center}
\includegraphics[width=90mm]{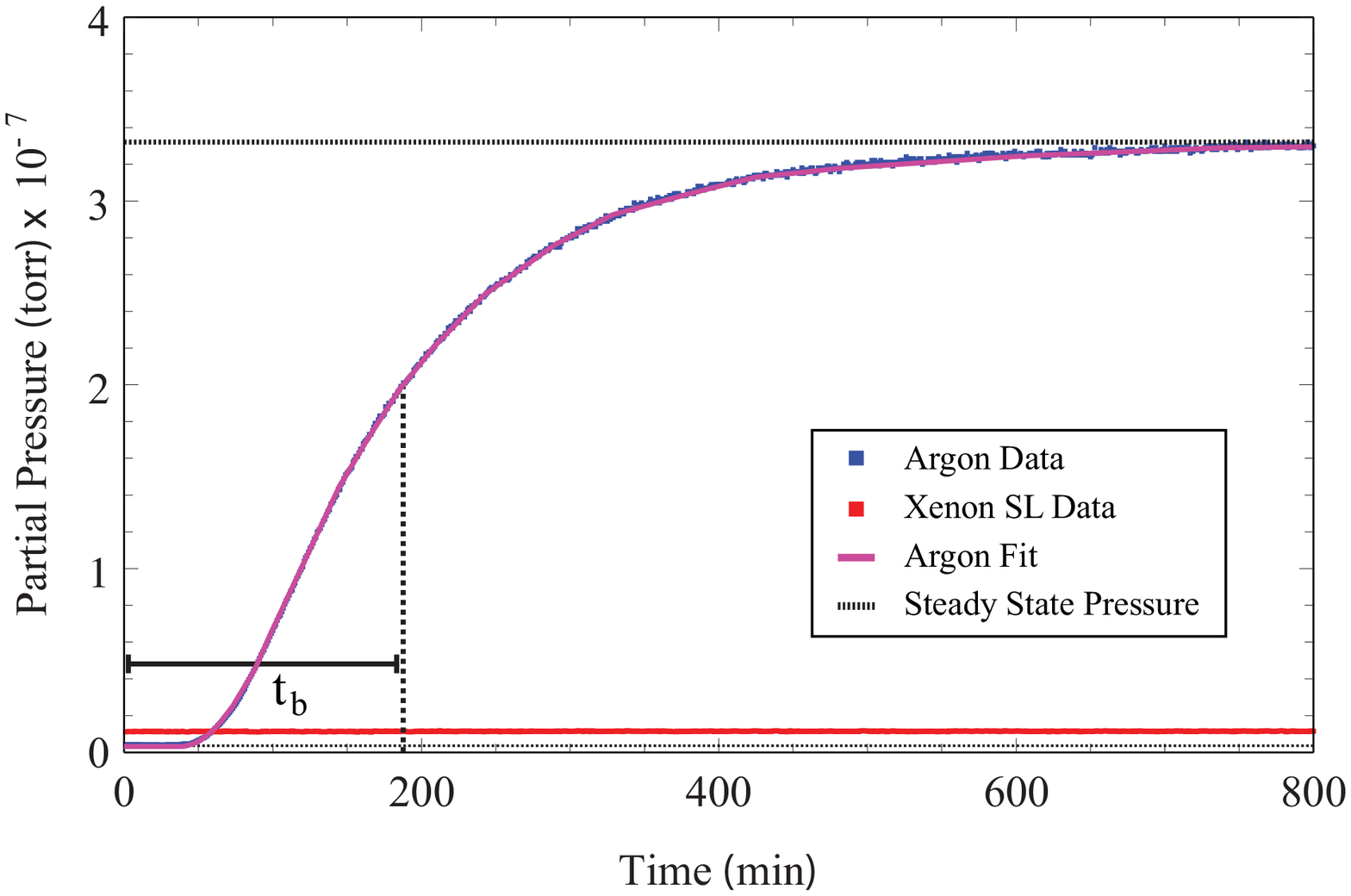}
\caption{(Color online) Sample data used to determine permeability and diffusivity. At $t=0$, gas is introduced to the high-pressure chamber and allowed to come in contact with the film.  Due to the pressure difference across the film, the gas begins to permeate the film. The argon gas partial pressure rises asymptotically to a steady state pressure after a characteristic breakthrough time, $t_b$, defined in Equation \ref{eq:lagtime}.}
\label{data}
\end{center}
\end{figure}

The time evolution of test gas partial pressure in the low-pressure chamber is analyzed to determine the permeability and the diffusivity of the test gas through the material under study. An example data run for argon permeating through Kapton is shown in Figure \ref{data}.
At $t=0$ argon gas is inserted into the high-pressure chamber and allowed to come in contact with the Kapton film. Argon diffuses through the film, causing the argon partial pressure in the low-pressure chamber to rise asymptotically to a steady state value, $P_{ss}$.  The diffusivity $D$ is determined by fitting the solution to the one-dimensional diffusion equation (Equation \ref{eq:diffeq}) to the partial pressure data shown in Figure \ref{data}. To fit this model to the data, we use terms up to $n = 3$, which provides less than 1\% deviation from the infinite sum over the entire fitting interval. The breakthrough time can then be calculated using Equation \ref{eq:lagtime}.  The fitting procedure is repeated for each experiment as the film material, film thickness, test gas, inlet pressure, and temperature are varied.

The test gas permeation rate $Q$ is determined by comparing the steady state pressure $P_{ss}$ to the steady state pressure $P_{SL}$ from the calibrated flow of the xenon standard leak, $Q_{SL}$. $P_{SL}$ was observed to remain unchanged for total pressures in the low-pressure chamber below $10^{-5}$ torr. $Q_{SL}$ can be expressed as $Q_{SL}=P_{SL}S_{\mathrm{eff}}^{Xe}$, where $S_{\mathrm{eff}}^{Xe}$ is the effective volumetric flow of xenon gas from the RGA to the pump.  Similarly, the flow rate of the permeating test gas can be written $Q_{gas}=P_{ss}S_{\mathrm{eff}}^{gas}=P_{ss}S_{\mathrm{eff}}^{Xe}\sqrt{m_{Xe}/m_{gas}}$, where the latter equality has used the linear dependence of volumetric flow on particle velocity in the molecular flow regime. Using Equation \ref{eq:permrate} we find that the permeability is given by
\begin{center}
\begin{equation}\label{eq:perm}
K=P_{ss}\frac{Q_{SL}}{P_{SL}} \sqrt{\frac{m_{Xe}}{m_{gas}}}\frac{d}{A\Delta P}.
\end{equation}
\end{center}
We can then use Equation \ref{eq:Db} to calculate the solubility $b$ from $K$ and $D$.

\begin{figure}[b]
\begin{center}
\includegraphics[width=90mm]{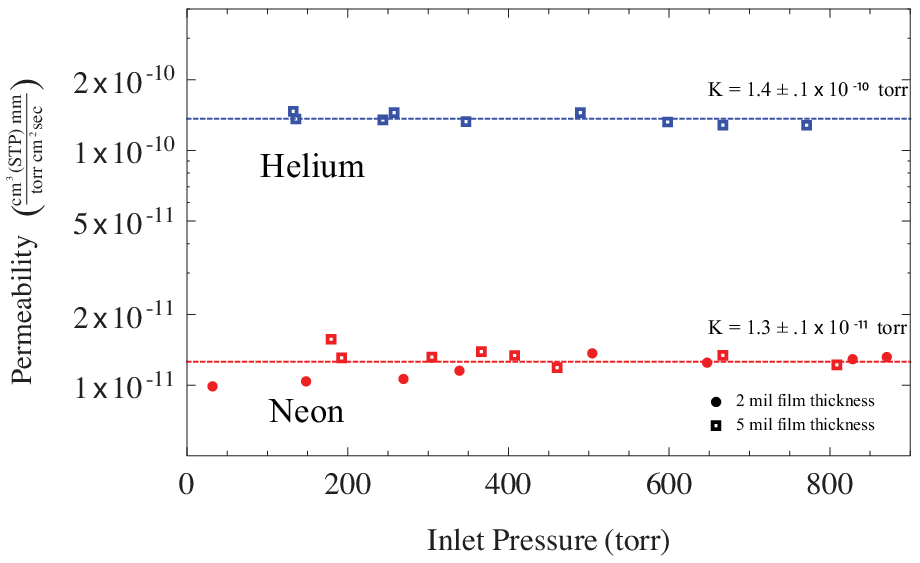}
\caption{(Color online) The permeability of Kapton is independent of film thickness and inlet pressure, as shown with He and Ne.}
\label{systematics}
\end{center}
\end{figure}

In order to check for systematic error, we varied several features of our experiment. To ensure that the test gas did not saturate the film material, we varied the inlet pressure of helium and neon and found that inlet pressure had no effect on $K$ (shown in Figure \ref{systematics}), implying that the film is not saturated over the test gas pressure range.  
We also tested the diffusive model of permeation by measuring $K$ and $D$ for neon permeating $2$ and $5$ mil thick Kapton films.  $K$ was unchanged by varying film thickness and $t_{b}$ increased by a factor of $6.2\pm .5$, consistent with the factor of $6.25$ predicted by the model for a constant $D$.  Lastly, we ensured $K$ and $D$ were not affected by varying the mesh size. Similar tests were repeated for each material studied.

As previously mentioned, the permeation rate can be manipulated by varying the temperature of the material.  Increasing the film temperature increases permeation rate, increasing $K$ and $D$ and decreasing $t_{b}$.  By measuring $K$ and $D$ at high temperatures, we can extrapolate room temperature data. Due to the properties of the materials, Kapton is the only material through which permeation at elevated temperatures were measured.

\begin{figure}[b]
\begin{center}
\includegraphics[width=90mm]{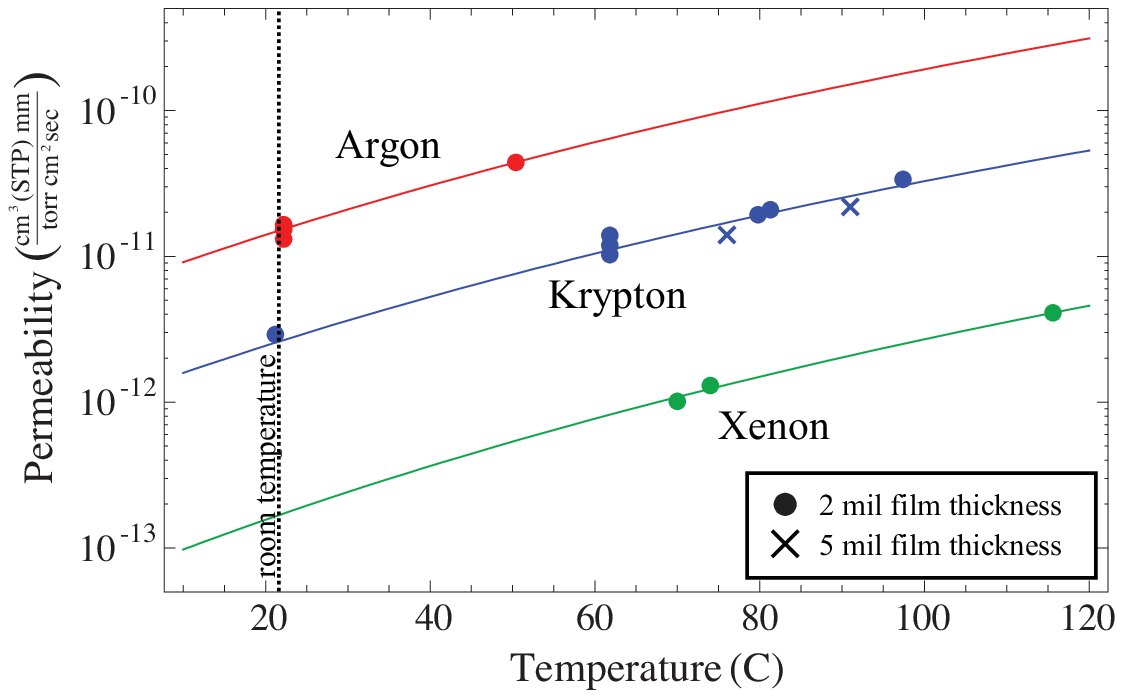}
\caption{(Color online) The temperature dependence of Ar, Kr, and Xe permeability through Kapton.  The curves are fits to Equation \ref{eq:temp_perm}.  The Xe fit is used to extrapolate the $22^{\circ}$C xenon permeability.}
\label{temp_perm}
\end{center}
\end{figure}

\begin{figure}[b]
\begin{center}
\includegraphics[width=90mm]{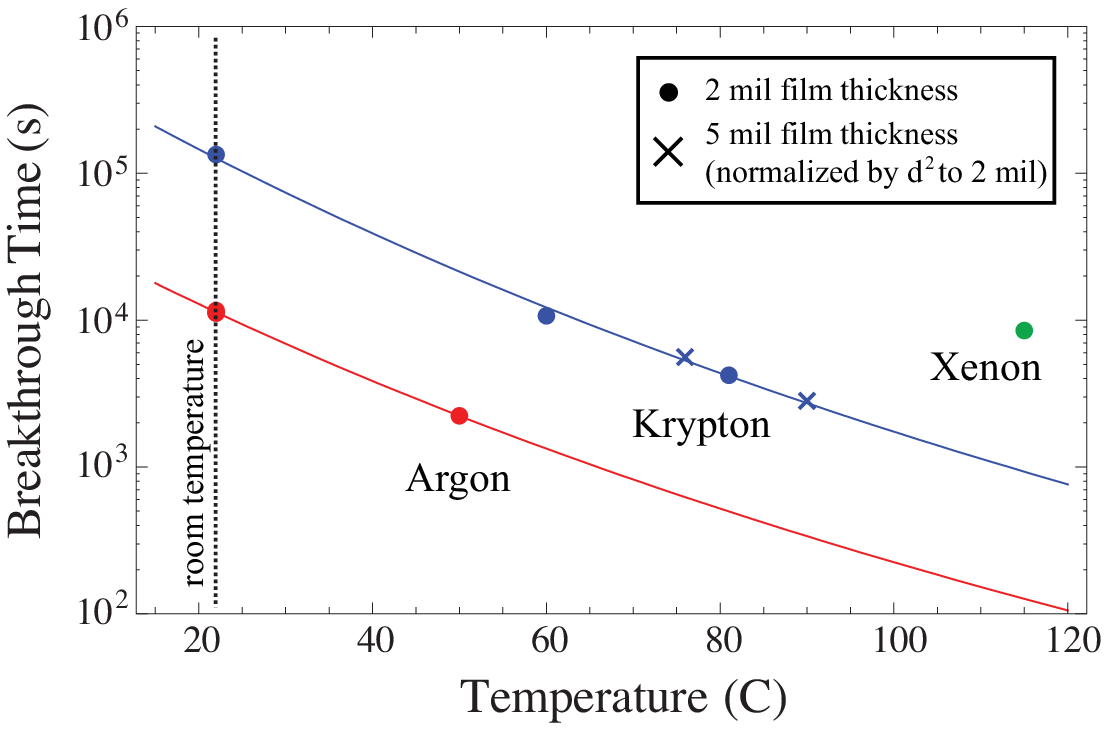}
\caption{(Color online) The temperature dependence of the breakthrough time of Ar and Kr permeating Kapton.  The data using 5 mil film thickness is scaled to 2 mil values for comparison using Equation \ref{eq:lagtime}.  The curves are fits to Equation \ref{eq:temp_tb}.}
\label{temp_tb}
\end{center}
\end{figure}

Using methods described above, we determined the permeability and diffusivity of argon, krypton, and xenon through $2$ and $5$ mil Kapton films at various temperatures.  These results are shown in Figures \ref{temp_perm} and \ref{temp_tb}.  For convenience, the diffusivities have been converted in Figure \ref{temp_tb} to breakthrough times through a $2$ mil film using Equation \ref{eq:lagtime}.
As expected, we observe an increase in $K$ and $D$ and thus a decrease in $t_{b}$ for each gas with increasing film temperature.  The Kapton film is not noticeably affected otherwise by the elevated temperatures, which are far below the melting point.

Measuring the permeability of xenon through Kapton at room temperature would take many days.  Instead, the data in Figure \ref{temp_perm} is fit to Equation \ref{eq:temp_perm} and we extrapolate the $22^{\circ}$C permeability of xenon through Kapton.  Room temperature breakthrough time of xenon through Kapton is not extrapolated due to insufficient $t_b$ data at high temperatures.

Using the methods discussed in the previous sections, we are able to determine the stable noble gas permeability, diffusivity, and solubility of the four materials studied at $22^{\circ}$C.  This data is shown in Table \ref{bigtable}.

\begin{table}[b]
  \begin{center}
\begin{tabular}{lcclll}
\hline
	\\[-10pt]
	Material	&Gas	&$K \mathrm{(\frac{cm^{3}~at~STP~mm}{s~torr~cm^{2}}}$)		&$D \mathrm{(\frac{cm^{2}}{s})}$ 	&$t_{b}$ (s) 	&$b (\mathrm{\frac{cm^{3}~at~STP}{torr~cm^{3}}})		$		\\[2pt]
			\hline
			\\[-4pt]
Kapton&He&$8.0\times10^{-10}$&$1.2\times10^{-6}$&$3.7$&$5.5\times10^{-4}$\\[4pt]
&Ne&$3.1\times10^{-11}$&$9.0\times10^{-8}$&$48$&$3.4\times10^{-5}$\\[4pt]
&Ar&$1.5\times10^{-11}$&$3.8\times10^{-10}$&$1.1\times10^{4}$&$3.9\times10^{-3}$\\[4pt]
&Kr&$2.9\times10^{-12}$&$3.2\times10^{-11}$&$1.3\times10^{5}$&$9.0\times10^{-3}$\\[4pt]
&Xe&$1.7\times10^{-13}~^{\dagger}$&&&\\
	\hline
			\\[-4pt]
Butyl&He&$1.0\times10^{-9}$&$9.5\times10^{-7}$&$4.5$&$1.1\times10^{-4}$\\[4pt]
&Ne&$7.4\times10^{-11}$&$2.0\times10^{-7}$&$22$&$3.8\times10^{-5}$\\[4pt]
&Ar&$1.8\times10^{-10}$&$1.9\times10^{-8}$&$2.3\times10^{2}$&$9.7\times10^{-4}$\\[4pt]
&Kr&$1.1\times10^{-10}$&$5.5\times10^{-9}$&$7.9\times10^{2}$&$2.0\times10^{-3}$\\[4pt]
&Xe&$2.7\times10^{-11}$&$3.7\times10^{-9}$&$1.2\times10^{3}$&$7.3\times10^{-4}$\\[4pt]
	\hline
			\\[-4pt]
Nylon&He&$1.8\times10^{-10}$&$7.3\times10^{-7}$&$5.9$&$2.5\times10^{-5}$\\[4pt]
&Ne&$7.4\times10^{-12}$&$9.2\times10^{-8}$&$47$&$8.0\times10^{-6}$\\[4pt]
&Ar&$5.4\times10^{-12}$&$1.0\times10^{-9}$&$4.3\times10^{3}$&$5.4\times10^{-4}$\\[4pt]
&Kr&$9.7\times10^{-13}$&$1.2\times10^{-10}$&$3.5\times10^{4}$&$7.9\times10^{-4}$\\[4pt]
&Xe&$6.3\times10^{-14}$&$7.4\times10^{-12}$&$5.8\times10^{5}$&$8.5\times10^{-4}$\\[4pt]
	\hline
			\\[-4pt]
Silver Shield&He&$6.9\times10^{-10}$&$1.9\times10^{-6}$&$2.2$&$3.6\times10^{-5}$\\[4pt]
&Ne&$2.1\times10^{-12}$&$1.4\times10^{-7}$&$30$&$1.5\times10^{-6}$\\[4pt]
&Ar&$2.5\times10^{-13}$&$4.2\times10^{-10}$&$1.0\times10^{4}$&$6.0\times10^{-5}$\\[4pt]
&Kr&$2.3\times10^{-14}$&$3.1\times10^{-11}$&$1.4\times10^{5}$&$7.5\times10^{-5}$\\[4pt]
	\hline
	\hline
	\\[-4pt]
Relative Uncertainty& &$50\%$&$10\%$&$10\%$&$50\%$\\[4pt]
\hline
\end{tabular}
\caption{Summary of room temperature permeation information for He, Ne, Ar, Kr, and Xe through Kapton, butyl, nylon, and Silver Shield. For convenient comparison, $t_b$ is calculated from Equation \ref{eq:lagtime} for $2$ mil material thickness. Uncertainty in $K$ and $b$ is based upon the systematic error in calibrating test gas flow with the xenon standard leak. Uncertainty in $D$ and $t_b$ is dominated by the uncertainty in determining film thickness. The value of $K$ for xenon permeating Kapton marked with a $\dagger$ has been extrapolated from higher temperature data using Equation \ref{eq:temp_perm}.}
\label{bigtable}
  \end{center}
\end{table}

\section{Model for Noble Gas Permeability of Polymers}\label{modelfor}

\begin{figure}[b]
\begin{center}
\includegraphics[width=90mm]{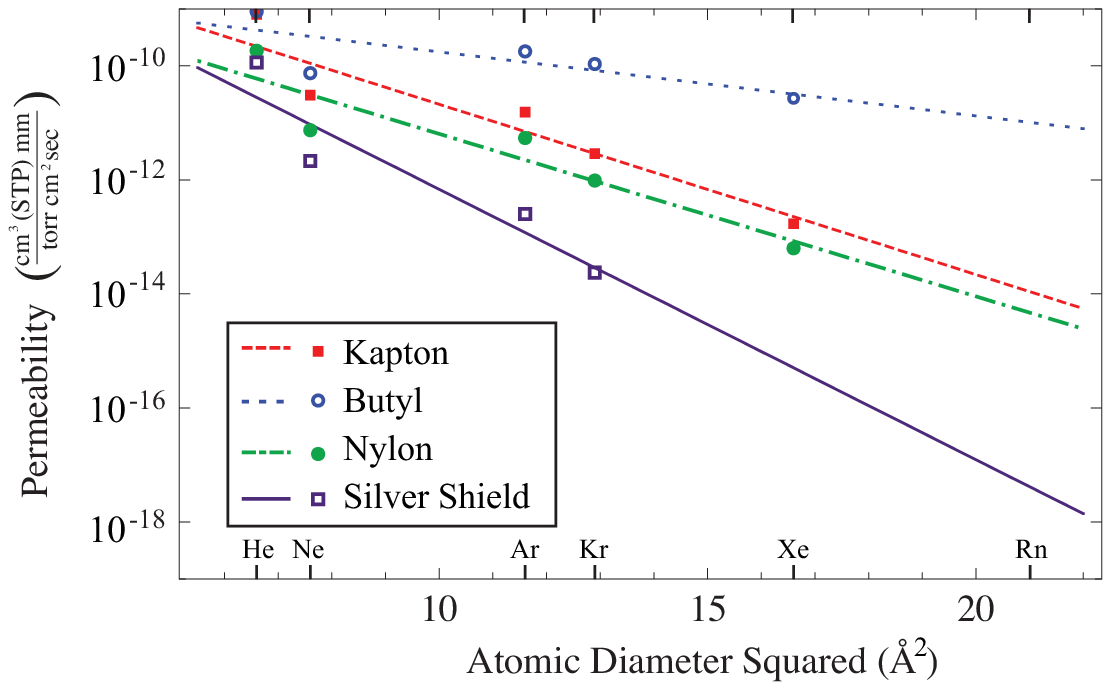}
\caption{(Color online) The exponential trend of room temperature ($22^{\circ}$C) permeabilities versus the square of the atomic diameter of the permeating gas.  Xenon permeation was not measured using Silver Shield.}
\label{model_perm}
\end{center}
\end{figure}

\begin{figure}[b]
\begin{center}
\includegraphics[width=90mm]{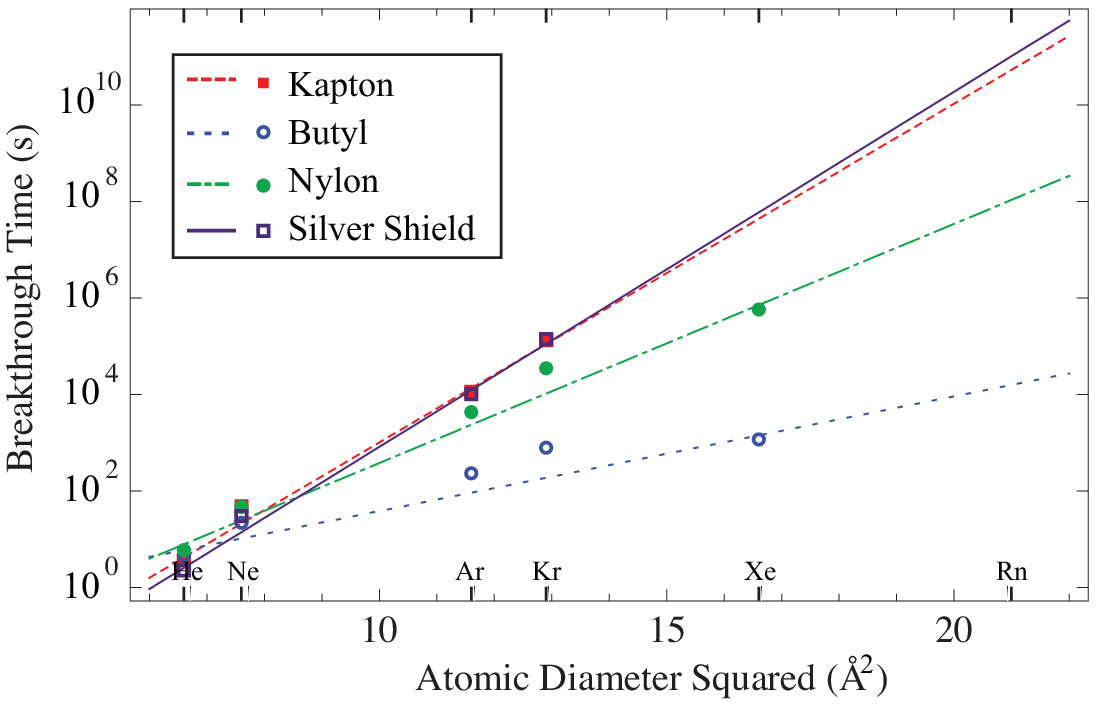}
\caption{(Color online) The exponential trend of room temperature ($22^{\circ}$C) breakthrough times versus the square of the atomic diameter of the permeating gas. The data are scaled to 2 mil thickness for comparison using Equation \ref{eq:lagtime}.  Xenon breakthrough time was not determined for Kapton or Silver Shield.}
\label{model_tb}
\end{center}
\end{figure}

The permeation of some polymers has been observed to show an exponential dependence with the square of the atomic radius of the permeating gas  \citep{hammon}. The noble gas permeabilities and breakthrough times of the four materials studied are plotted in this manner in Figures \ref{model_perm} and \ref{model_tb} along with exponential fits for each material.  The atomic radii, taken from \citep{hirschfelder}, are the same as those used in \citep{hammon}.

The room temperature permeation of xenon through Silver Shield is not measured due to the length of time required for the measurement.  Silver Shield material cannot be heated to temperatures above $50^{\circ}$C, thus we cannot decrease the experimental time and extrapolate room temperature values in the manner described above.  Room temperature diffusivity of xenon though Kapton is also not included due to insufficient data for extrapolation.

Using the empirical model described above, we estimate $K$ and $t_{b}$ for radon permeation through Kapton, butyl, nylon, and Silver Shield at $22^{\circ}$C.  These estimates are shown in Table \ref{rntable}.
\begin{table}[b]
  \begin{center}
    \begin{tabular}{llll}
	\hline
	\\[-10pt]
	Material		&	Value	&	Rn Estimation		&	Xe Bound		\\[2pt]
	\hline
			\\[-4pt]
	Kapton		&	$K$		&	$1^{}_{}\times10^{-14} $	&     $1.7\pm0.8\times10^{-13}$				 \\[4pt]
				&	$t_{b}$	&	$5^{}_{}\times10^{10} $	&    $ 1.3\pm0.1\times10^{5}$ (Kr Bound)				\\[4pt]
			\hline	
			\\[-4pt]
	Butyl			&	$K$		&	$1^{}_{}\times10^{-11}$	&   $  2.7\pm1.4\times10^{-11}$				\\[4pt]
				&	$t_{b}$	&	$2^{}_{}\times10^{4}$	&    $ 1.2\pm0.1\times10^{3}	$			\\[4pt]
			\hline
			\\[-4pt]
	Nylon		&	$K$		&	$5^{}_{}\times10^{-15}$	&    $  6.3\pm3.2\times10^{-14}$				\\[4pt]
				&	$t_{b}$	&	$1^{}_{}\times10^{8}$	&    $  5.8\pm0.6\times10^{5}	$			\\[4pt]
			\hline
			\\[-4pt]
	Silver Shield	&	$K$		&	$4^{}_{}\times10^{-18}$	&     $  2.3\pm1.2\times10^{-14}$ (Kr Bound)	\\[4pt]
				&	$t_{b}$	&	$1^{}_{}\times10^{11}$	&    $  1.4\pm0.1\times10^{5}$ (Kr Bound)		\\[4pt]
	\hline
	\newline
	\end{tabular}
\caption{Summary of estimations and bounds for room temperature Rn $K$ and $t_b$ through $2$ mil material.  The uncertainties for the Xe and Kr bounds are the same as in Table \ref{bigtable}.  The Xe and Kr bounds reflect the systematic uncertainty of the model used to estimate the Rn values.}
\label{rntable}   
  \end{center}
\end{table}
Both $K$ and $t_{b}$ are typically monotonic with respect to the square of the atomic diameter of the permeating gas.  Thus the measured values for xenon can be taken as a conservative upper and lower bound for the radon values for $K$ and $t_{b}$ respectively.  These bounds are also included in Table \ref{rntable}.  Krypton bounds are used for materials whose xenon permeation values were not measured.   

The uncertainty of the radon permeation estimates is dominated by systematic uncertainty in applying the model function.  Although the fits of permeability to this function in Figure \ref{model_perm} agree with the data rather well over several orders of magnitude of permeability, and similar fits in \citep{hammon} provided realistic estimates of radon permeability, there is considerable uncertainty in the extrapolations to radon.  For example, the fit underestimtes helium permeability and overestimates neon permeability for all materials studied, suggestive of a more complex functional form.  Similarly, the fits of breakthrough times to the same model in Figure \ref{model_tb} assume a similar or weak dependence of solubility on the square of the atomic diameter, which may not be the case.  With the limitations of the model in mind, the bounds given by xenon or krypton measurements reflect the estimation uncertainty.

The radon isotope of concern to low radioactive background experiments is radon-$222$, which has a half-life of $3.8$ days ($3.3\times10^{5}$ s) \citep{ramachandran}.  A gasket suitably impermeable to radon for these experiments should have a breakthrough time that is long compared to the radon-$222$ half-life.  Since $t_b \propto d^2$, $t_b$ can be greatly increased by increasing the distance over which gas permeates.  If $t_b$ is much longer than the radon-$222$ half-life, then only a small fraction of radon atoms will permeate a gasket before decaying. Additionally, the radon exposure time can be minimized to reduce the total number of dissolved radon atoms.

\section{Conclusion}
We use a gas-flow method to measure and calculate previously unrecorded data for noble gas permeability, diffusivity, and solubility of Kapton, butyl, nylon, and Silver Shield at $22^{\circ}$C.  We note that these properties can vary on the details of manufacture and especially between different manufacturers.  The temperature dependence of permeation can be exploited to manipulate the permeation rate, as demonstrated here. The permeability of Kapton is measured at higher temperatures up to $120^{\circ}$C using argon, krypton, and xenon, and these values are used to extrapolate the xenon permeability of Kapton at $22^{\circ}$C. Based on the empirical model used previously in \citep{hammon}, we estimate radon permeability and breakthrough time of $2$ mil films at $22^{\circ}$C.  With this information, the suitability of the use of the materials studied as gasket or glove materials in low background radiation experiments can be appropriately determined.

\section{Acknowledgments}
We would like to thank the DEAP/CLEAN collaboration for helpful discussions, and the Weak Interactions team at Los Alamos National Laboratory for the suggestion of testing Silver Shield and for providing the nylon material.

\bibliographystyle{unsrtnat}
\bibliography{permeation_NIMA}

\end{document}